\documentclass[aps, prd, 10pt, twocolumn, nofootinbib, superscriptaddress, elide, preprintnumbers]{revtex4-1}
\pdfoutput=1
\usepackage[T1]{fontenc}
\usepackage[utf8]{inputenc}
\usepackage{lmodern, amsmath, amsfonts, feynmp, slashed, multirow,graphicx}
\usepackage[colorlinks=true, citecolor=blue, urlcolor=blue, linkcolor=blue, breaklinks=true, pdfpagelabels=false]{hyperref}

\makeatletter
\g@addto@macro\bfseries{\boldmath}
\makeatother
\allowdisplaybreaks[4]

\def\equationautorefname~#1\null{Eq.\,(#1)\null}
\def\pageautorefname\nobreakspace{p.}

\makeatletter%
\renewcommand{\p@subsection}{\thesection.}
\makeatother%
\graphicspath{{fig/}}
\newcommand{\myscale}{.88}

\newcommand{\sym}[1]{\textrm{#1}}
\renewcommand{\P}{\sym{P}}
\newcommand{\M  }{{\ensuremath{\hat{\sym{T}}}}}

\newcommand{\CP }{\sym{CP}}
\newcommand{\T  }{\sym{T}}

\newcommand{\E  }{\sym{E}}
\newcommand{\ps}{\Phi}

\newcommand{\tp}[1][\empty]{\emph{triple\ifx#1\empty\ product\else-product\fi}}

\DeclareMathAlphabet{\mathsf}{\encodingdefault}{\sfdefault}{m}{sl}

\DeclareMathOperator{\Br}{Br}

\DeclareMathOperator{\sig}{sign}
\let\Re\undefined \DeclareMathOperator{\Re}{Re}
\let\Im\undefined \DeclareMathOperator{\Im}{Im}
\renewcommand{\d}{\text{d}}
\renewcommand{\v}{\boldsymbol}

\makeatletter
\newcommand{\parenbar}{\mathpalette\p@renb@r}
\def\p@renb@r#1#2{%
	\vbox{%
		\ifx#1\scriptscriptstyle\dimen@.7em\dimen@ii.20em\else%
		\ifx#1\scriptstyle	\dimen@.8em\dimen@ii.25em\else%
					\dimen@.9em\dimen@ii.35em\fi\fi%
		\offinterlineskip%
		\ialign{%
			\hfill##\hfill\cr
			\vbox{\hrule width\dimen@ii height .32pt}\cr
			\noalign{\vskip-.35ex}%
			\hbox to\dimen@{$\mathchar300\hfil\mathchar301$}\cr
			\noalign{\vskip-.35ex}%
			$#1#2$\cr%
		}%
	}%
}%
\makeatother
\newcommand{\parenpm}{%
	\raisebox{-.8ex}{$\mathchar300$}%
	\ensuremath{\hspace{-.7ex}\pm\hspace{-.7ex}}%
	\raisebox{-.8ex}{$\mathchar301$}}

\newcommand{\dgs}[2]{\left.\frac{\d\Gamma}{\d\ps}\right| 
	\ifx#1\empty\else^{\text{#1}}\fi%
	\ifx#2\empty\else_{\text{#2}}\fi}
\newcommand{\ips}{\int\hspace{-.75ex}\d\ps\hspace{.5ex}} 
\newcommand{\ct}[2][]{\,\text{c}\ifx#1\empty\else#1\fi\theta%
	\ifx#2\empty\else_%
		\ifx#21 a\else\ifx#22 b\else #2\fi\fi\fi\,}
\newcommand{\st}[2][]{\,\text{s}\ifx#1\empty\else#1\fi\theta%
	\ifx#2\empty\else_%
		\ifx#21 a\else\ifx#22 b\else #2\fi\fi\fi\,}
\newcommand{\cp}[1][]{\,\text{c}\ifx#1\empty\else#1\fi\phi\,}
\renewcommand{\sp}[1][]{\,\text{s}\ifx#1\empty\else#1\fi\phi\,}
\newcommand{\pa}{\parallel}
\newcommand{\pe}{\perp}

\begin{document}

\setlength{\unitlength}{1mm}
\begin{fmffile}{graphs}
\fmfcmd{%
prologues:=3;
thin := .5pt;
arrow_ang := 20;
arrow_len := 3.5thick;
curly_len := 2.5thick;
color mb, mr, mg; mb :=.5blue; mr :=.5red; mg :=.5green;
style_def top expr p =
  save oldpen;
  pen oldpen;
  oldpen := currentpen;
  pickup oldpen scaled 3.5;
  ccutdraw p;
  pickup oldpen scaled 1.5;
  cullit; undraw p; cullit;
  cfill (arrow p);
enddef;
}

\title{Probing \CP\ violation systematically\texorpdfstring{\\}{ }in differential distributions}

\newcommand{\cornell}{Laboratory for Elementary Particle Physics, Cornell University, Ithaca, NY 14853, USA}
\author{Gauthier Durieux}
\affiliation{\cornell}
\affiliation{Centre for Cosmology, Particle Physics and Phenomenology, 
Université catholique de Louvain, B-1348 Louvain-la-Neuve, Belgium}
\author{Yuval Grossman}
\affiliation{\cornell}

\preprint{CP3-15-24}

\begin{abstract}
We revisit the topic of \tp[-] asymmetries which probe \CP\ violation through
differential distributions. We construct distributions with well-defined discrete
symmetry properties and characterize the asymmetries formed upon them. It is
stressed that the simplest asymmetries may not be optimal. We explore systematic
generalizations having limited reliance on the process dynamics and phase-space
parametrization. They exploit larger fractions of the information contained in
differential distributions and may lead to increased sensitivities to \CP\
violation. Our detailed treatment of the case of spinless four-body decays
paves the way for further experimental studies.
\end{abstract}

\maketitle

\section{Introduction}

The fully differential rates of some multibody meson decays are being more and
more accurately measured. In the search for new sources of \CP\ violation, such
processes present several advantages. They often feature a rich variety of
interfering contributions from which differences in
\CP-violating---\emph{weak}---phases could manifest themselves. In addition, the
multiplication of measurable independent four-vectors permits the construction of
so-called \tp[-] observables. These have a couple of interesting characteristics.
Unlike total rate asymmetries between \CP-conjugate processes, their sensitivity
to small differences in \CP-violating phases is not conditioned by the presence
of \CP-conserving---\emph{strong} or \emph{unitary}---phase differences. They
can also be measured using \emph{untagged} samples in which \CP-conjugate
processes need not be distinguished, provided their fractions are equal.

In this paper, we explore the variety of possible \tp[-]
observables. The ever-increasing amount of data collected allows finer details
of the differential distributions for which they are proxies 
to become measurable. We stress that the most common asymmetries may not be the most sensitive
ones, due to cancellations in phase-space integrals. As much as possible, we
would like to abstract our treatment from the particular dynamics of the studied
process. In many multibody decays, only phenomenological descriptions
of various degrees of accuracy are achieved. They may not capture all the fine
details of interfering contributions which could reveal \CP\ violation. A
systematic procedure that is less likely to miss unpredicted forms of \CP\
violation is therefore desirable. Although we will mostly focus, for
concreteness, on four-body meson decays involving spinless particles, our
discussion has a wider range of application.

\subsection{Differential \CP\ violation}
Let us consider two transitions of amplitudes $\mathcal{M}(\{\lambda_i,p_i\})$ and
$\bar{\mathcal{M}}(\{\lambda_{\bar\imath},p_{\bar\imath}\})$. They involve an equal number particles respectively labeled by $i$ and $\bar\imath$, with helicities $\lambda_{i,\bar\imath}$ and
four-momenta $p_{i,\bar\imath}$. We would like to
perform a comparison of these two amplitudes phase-space point by phase-space
point so we take $\lambda_{\bar\imath}=\lambda_i$ as well as
$p_{\bar\imath}=p_{i}$.

If these two processes are \CP\ conjugate of each other, with $\bar\imath=\CP[i]$, \CP\ violation at any
phase-space point takes the form of a difference between the squared moduli
of
\begin{equation*}
	\mathcal{M}(\{\lambda_i,p_i\})
	\quad\text{and}\quad
	\bar{\mathcal{M}}(\{\lambda_i,\bar p_i\})
\end{equation*}
where $\bar p\equiv \P[p]$ is the parity conjugate of the momentum $p$. Testing
\CP\ conservation phase-space point by phase-space point thus implies a
comparison of the differential rates of two processes involving \CP-conjugate
particles of identical helicities but opposite three-momenta.

It reveals useful to define an operator, called \emph{motion reversal} and
denoted here by \M, that reverts both momentum and spin
three-vectors~\cite{Sachs:1987gp, Branco:1999fs}. Its action on helicities and
momenta is thus identical to that of \CP\ and it can be viewed as the unitary
component of the antiunitary time-reversal operator \T. It is therefore
sometimes called \emph{naive \T}. In general, the amplitudes above can then be
decomposed into two pieces that are respectively \M-even and
\M-odd~\cite{Nowakowski:1989ju,*Nowakowski:1988dx}:
\newcommand{\me}{{\mathcal{M}_e}}
\newcommand{\mo}{{\mathcal{M}_o}}
\newcommand{\lp}[1][]{(\{\lambda_i,#1p_i\})}
\begin{align*}
\mathcal{M}\lp &=
	 \me\lp
	+\mo\lp,
\\[2mm]
\bar{\mathcal{M}}\lp[\bar] &=
	 \bar\me\lp[\bar]
	+\bar\mo\lp[\bar]
\\&=
	 \bar\me\lp
	-\bar\mo\lp.
\end{align*}
Those two terms can receive several contributions whose absorptive
parts~\cite{Watson:1952ji, Slobodrian:1971an} take the form of \CP-even phases
$\delta$. One
can then write
\begin{equation}
\begin{array}{l}
		\me\lp
	 	= a_e^j \; e^{i(\delta_e^j+\varphi_e^j)},
	\\[1mm]
	\bar	\me\lp
		= a_e^j \; e^{i(\delta_e^j-\varphi_e^j)},
	\\[3mm]
		\mo\lp
		= a_o^k \; e^{i(\delta_o^k+[\varphi_o^k+\pi/2])},
	\\[1mm]
	\bar	\mo\lp
		= a_o^k \; e^{i(\delta_o^k-[\varphi_o^k+\pi/2])},
\end{array}
\label{eq:ifactor}
\end{equation}
with implicit summation over the $j,k$ indices, and real $a_{e,o}^{j,k}$,
$\delta_{e,o}^{j,k}$, $\varphi_{e,o}^{j,k}$ functions of the helicities and
momenta $\{\lambda_i,p_i\}$. The above conventions imply that all \CP\ violation
is encoded in the \CP-odd phases $\varphi_{e,o}^{j,k}$. When they vanish,
\begin{align*}
	&\bar\me\lp = +\me\lp,\\
	&\bar\mo\lp = -\mo\lp,
\end{align*}
so that the \CP-conjugate rates are identical, phase-space point by phase-space
point. As the physical amplitude is defined up to an overall phase, a departure
from zero for differences in these $\varphi_{e,o}^{j,k}$ is what we are after.

\subsection{\CP\ violation without \CP-even phases}
\label{sec:phases}
The \M-transformed differential rates are obviously accessible experimentally
since the measured momenta can be artificially reversed. For processes involving
only scalars in their initial and final states, \M\ is actually equivalent to
parity conjugation \P. The measured differential rates of any pair of
\CP-conjugate processes can therefore be decomposed into four pieces of definite
\M\ and \CP\ transformation properties:
\begin{equation}
\dgs	{\M-$ _\text{odd}^\text{even}$}
	{\CP-$_\text{odd}^\text{even}$}
\equiv
\frac{\mathbb{I} \pm \M}{2}\; 
\frac{\mathbb{I} \pm \CP}{2}
\;\frac{\d\Gamma}{\d\ps}
\label{eq:decomp}
\end{equation}
with the shorthand $\ps\equiv\{\lambda_i,p_i\}$.

For simplicity, let us assume there are respectively two and one contribution(s)
to the \M-even and \M-odd parts of the amplitude in the process under scrutiny:
\begin{equation*}
	\begin{array}{*{3}{r@{\,}l@{}l}}
	\mathcal{M}(\{\lambda_i,p_i\}) =
			& a_e^1 & e^{i(\delta_a^1+\varphi_a^1)}
		+\!\!	& a_e^2 & e^{i(\delta_a^2+\varphi_a^2)}
		+i	& a_o^1\! & e^{i(\delta_o^1+\varphi_o^1)},\\[1mm]
	\bar{\mathcal{M}}(\{\lambda_i,\bar p_i\}) =
			& a_e^1 & e^{i(\delta_a^1-\varphi_a^1)}
		+\!\!	& a_e^2 & e^{i(\delta_a^2-\varphi_a^2)}
		+i	& a_o^1\! & e^{i(\delta_o^1-\varphi_o^1)}.
	\end{array}
\end{equation*}
All functions of the phase space are evaluated at $\{\lambda_i,p_i\}$.
Note the convention of \autoref{eq:ifactor} causes the
appearance of a factor of $i$ in front of the \M-odd term.
Up to a flux factor, the squared modulus of this expression and of its \CP\
conjugate provides us with the differential rates which can be decomposed as
prescribed in \autoref{eq:decomp}:
\newcommand{\eone}{a_e^1}
\newcommand{\etwo}{a_e^2}
\newcommand{\oone}{a_o^1}
\begin{equation*}
\begin{array}{r @{\:}r @{\;\:}l @{\:}l}
\displaystyle\dgs{\M-even}{\CP-even}\propto
  &	\multicolumn{3}{l}{
	  \eone\;\eone
	+ \etwo\;\etwo
	+ \oone\;\oone
	}
\\&
	+2\,\eone \; \etwo
		&\cos(\delta_e^1-\delta_e^2)	&\cos(\varphi_e^1-\varphi_e^2),
\\[1mm]
\displaystyle\dgs{\M-odd}{\CP-even}\propto
  &	 2\,\eone \; \oone
		&\sin(\delta_e^1-\delta_o^1)	&\cos(\varphi_e^1-\varphi_o^1)
\\&	+2\,\etwo \; \oone
		&\sin(\delta_e^2-\delta_o^1)	&\cos(\varphi_e^2-\varphi_o^1),
\\[1mm]
\displaystyle\dgs{\M-even}{\CP-odd}	\propto
  &	-2\,\eone \; \etwo
		&\sin(\delta_e^1-\delta_e^2)	&\sin(\varphi_e^1-\varphi_e^2),
\\[5mm]
\displaystyle\dgs{\M-odd}{\CP-odd}	\propto
  &
	 2\,\eone \; \oone
		&\cos(\delta_e^1-\delta_o^1)	&\sin(\varphi_e^1-\varphi_o^1)
\\&	+2\,\etwo \; \oone
		&\cos(\delta_e^2-\delta_o^1)	&\sin(\varphi_e^2-\varphi_o^1).
\end{array}
\end{equation*}
The last two expressions above vanish in the CP limit. There are thus
two distinct kinds of \CP-violating differential
rates~\cite{Gasiorowicz:1966xra}: the presence of the \M-even one requires
nonvanishing differences in \CP-even phases $\delta$ while the
\M-odd--\CP-odd does not. This can be understood as, in the absence of
absorptive part to the amplitude, \M\ is equivalent to \T\ so that \CP\T\
conservation imposes any \CP-odd quantity to also be \M\ odd~\cite{Rindani:1994ad}.

On the other hand, the \M-odd--\CP-even piece of the differential rate could be
used to isolate relatively small differences in \CP-even phases $\delta$, in the
absence of \CP-odd phase $\varphi$. It can thus serve to better understand
final-state interactions.

\subsection{Untagged samples}
\label{sec:untagged}
Another remarkable characteristic of the \M-odd-\CP-odd part of the differential
rate is that it can be measured with samples which contain an equal number of
events from \CP-conjugated processes. It can also be evaluated in the decay of
self-conjugate states like the $Z$ and $h$ bosons, or any Majorana fermion. This
can be understood by rewriting the \M-odd--\CP-odd differential rate defined in
\autoref{eq:decomp} as
\begin{equation*}
	\frac{\mathbb{I} - \M }{2}\;
	\bigg(
	\frac{\mathbb{I} + \CP\M }{2}
	\;\frac{\d\Gamma}{\d\ps}
	\bigg),
\end{equation*}%
using the fact that \M\ is an involution: $\M^2=\mathbb{I}$. It only involves
$\d(\Gamma+\bar\Gamma)/\d\ps$ evaluated at the phase-space point
$\{\lambda_i,p_i\}$ and at its \M\ conjugate $\{\lambda_i,\bar p_i\}$.

Other discrete symmetry operators can be introduced. In particular, let us
denote a permutation of the external particles as $\E\{i_1,i_2,\ldots,i_n\} =
\{\E[i_1],\E[i_2],\ldots,\E[i_n]\}$. For transitions involving a self-conjugate
subset of external particles, there is an especially relevant permutation $\E^*$
that takes each particle in the subset to its \CP\ conjugate. For example,
$\E^*\{K^+, K^-, \pi^+,\pi^-\} = \{K^-,K^+,\pi^-,\pi^+\}$.

A part of the differential rate that is odd under a permutation \E\ can also be
used to test \CP\ conservation with samples containing an equal number of
events from \CP-conjugate processes:
\begin{equation*}
	\frac{\mathbb{I} - \E }{2}\;
	\bigg(
	\frac{\mathbb{I} + \CP\,\E }{2}
	\;\frac{\d\Gamma}{\d\ps}
	\bigg).
\end{equation*}
However, resorting to such samples is only desirable when a subset of the
particles involved is self-conjugate. Experimentally, the tagging that
discriminates between the \CP-conjugate processes then comes with
an efficiency cost. Importantly, without tagging, what is then
actually measured is
\begin{equation*}
	\frac{\mathbb{I} + \CP\M\,\E^*}{2}
	\;\frac{\d\Gamma}{\d\ps}.
\end{equation*} 
In an untagged sample, one can therefore measure two \CP-odd
differential rates that are either \M-odd--$\E^*$-even or \M-even--$\E^*$-odd:
\begin{multline*}
	\frac{\mathbb{I} \pm \M}{2}\;
	\frac{\mathbb{I} \mp \E^*}{2}\;
	\frac{\mathbb{I} - \CP}{2}
	\;\frac{\d\Gamma}{\d\ps}
	\\=
	\frac{\mathbb{I} \pm \M}{2}\;
	\frac{\mathbb{I} \mp \E^*}{2}\;
	\bigg(
	\frac{\mathbb{I} + \CP\M\,\E^*}{2}
	\;\frac{\d\Gamma}{\d\ps}
	\bigg).
\end{multline*}
Some asymmetries of either kind were measured by the LHCb Collaboration in its
study of the $B^0_s\to K^+K^-\pi^+\pi^-$ decay with an untagged
sample~\cite{Aaij:2015kba} (see discussion in \autoref{sec:untagged_ex}).

On the contrary, the differential rates of identical \M\ and $\E^*$ parities are
\CP\ even in an untagged sample:
\begin{multline*}
	\frac{\mathbb{I} \pm \M}{2}\;
	\frac{\mathbb{I} \pm \E^*}{2}\;
	\bigg(
	\frac{\mathbb{I} + \CP\M\,\E^*}{2}
	\;\frac{\d\Gamma}{\d\ps}
	\bigg)
	\\=
	\frac{\mathbb{I} \pm \M}{2}\;
	\frac{\mathbb{I} \pm \E^*}{2}\;
	\frac{\mathbb{I} + \CP}{2}
	\;\frac{\d\Gamma}{\d\ps}
	.
\end{multline*}
As in the tagged sample case, they provide a handle on the \CP-even phases.

\subsection{Integrated observables}
\label{sec:int_obs}
No phase-space integration or spin averaging is in principle required to test
for the existence of \CP-violating phases. Such procedures are only applied
because of practical constraints like finite statistics. The total
rate asymmetry is constructed upon the \M-even--\CP-odd differential rate
\begin{equation}
	\ips\dgs{\M-even}{\CP-odd}
	\label{eq:rate_asym}\,.
\end{equation}
A second family of observables can be obtained from integrals of its
\M-odd--\CP-odd homologue
\begin{equation}
	\ips\:\mathsf{f}(\ps)\:\dgs{\M-odd}{\CP-odd}
	\label{eq:tp_int}
\end{equation}
with some \M-odd function $\mathsf{f}(\ps)$ without which the phase-space
integral would vanish. Similarly, any \M-even function $\mathsf{g}(\ps)$ could
be inserted into the \M-even--\CP-odd integral to construct observables sharing
the properties of the total rate asymmetry.

As a product of a \M-odd kinematic function with a \M-odd--\CP-even differential
rate, the observables of \autoref{eq:tp_int} are \M\ even and \CP\ odd but have
not definite \T\ transformation properties.

\subsection{\texorpdfstring{\M}{T} oddity and triple products}
There are two tensors available to construct Lorentz invariants from spin and
momenta four-vectors. The metric $g_{\mu\nu}$ leads to \M-even contractions like
invariant masses, and the completely antisymmetric $\epsilon_{\mu\nu\rho\sigma}$
produces \M-odd combinations of four-vectors.

Dot products and antisymmetric contractions of four-momenta and Pauli-Lubański
spin vectors (respectively denoted by $p$ and $w$) have definite \P\
parities. The \P-even combinations are:
\vspace*{-2mm}
\begin{gather*}
	p_1\cdot p_2,
	\quad
	w_1\cdot w_2,
	\\
	\epsilon_{\mu\nu\rho\sigma}\; p_1^\mu p_2^\nu p_3^\sigma w_4^\rho,
	\quad\text{and}\quad
	\epsilon_{\mu\nu\rho\sigma}\; p_1^\mu w_2^\nu w_3^\sigma w_4^\rho,
\end{gather*}
while the \P-odd ones are:
\vspace*{-1mm}
\begin{gather*}
	p_1\cdot w_2,
	\\
	\epsilon_{\mu\nu\rho\sigma}\; p_1^\mu p_2^\nu p_3^\sigma p_4^\rho,\quad
	\epsilon_{\mu\nu\rho\sigma}\; p_1^\mu p_2^\nu w_3^\sigma w_4^\rho,\quad
	\epsilon_{\mu\nu\rho\sigma}\; w_1^\mu w_2^\nu w_3^\sigma w_4^\rho.
\end{gather*}
The sensitivities to discrete symmetry violation of observables
having definite \P\ and \M\ transformation properties, in the presence or
absence of absorptive parts in the amplitude, are listed on p.~519 of
Ref.~\cite{Gasiorowicz:1966xra}.

The completely antisymmetric Lorentz structure can originate
directly from Lagrangian couplings like $i\epsilon_{\mu\nu\rho\sigma} F^{\mu\nu}
F^{\rho\sigma}$, or arise in the presence of chiral fermions, since $\gamma^5 =
\frac{i}{4!}\: \epsilon_{\mu\nu\rho\sigma} \gamma^\mu \gamma^\nu \gamma^\rho
\gamma^\sigma$. Because it is completely antisymmetric, however, a necessary
condition for the presence of a \M-odd part $\mo$ in an amplitude is the
availability of four independent and distinguishable four-vectors. In a process
involving scalars or particles of unmeasured spins, at least five external momenta are
therefore required.

In a reference frame where $a^\mu=(a^0,\v{0})$, the completely antisymmetric
combination of four four-vectors $\epsilon_{\mu\nu\rho\sigma} \; a^\mu \, b^\nu
\, c^\rho \, d^\sigma$ reduces to a $a^0\; \v{b}\cdot (\v{c}\times\v{d})$ scalar
\tp\ (for $\epsilon_{0123}\equiv+1$). The observables constructed from the
\M-odd parts of the differential rate are therefore customarily called \tp[-]
asymmetries. A significant amount of effort, both theoretical and experimental
has been devoted to their study. A \tp[-] asymmetry has been measured in $K_L\to
\pi^+\pi^- e^+e^-$ \cite{AlaviHarati:1999ff,*Lai:2003ad,*Abouzaid:2005te} and
applications are also found in heavy-meson~\cite{Dell'Aquila:1985ve,
*Dell'Aquila:1985vc, *Dell'Aquila:1985vb, Donoghue:1987wu, Valencia:1988it,
Kayser:1989vw, Dunietz:1990cj,Kramer:1991xw, *Kramer:1991ab, *Kramer:1992gi,
*Kramer:1993yu, Atwood:1994kn, Kruger:1999xa, Bensalem:2000hq, Datta:2003mj,
Egede:2008uy, Kang:2009iy, Gronau:2011cf, Datta:2011qz, Atwood:2012ac,
Descotes-Genon:2013vna, Bevan:2014nva}~\cite{Aubert:2008zza, Aaij:2014qwa,
Aaltonen:2011rs, *Aaij:2012ud, *Aaij:2014kxa, Aaij:2013swa, Aaij:2014tpa},
baryon~\cite{Kang:2010td,Gronau:2015gha}, top~\cite{Atwood:2000tu},
$Z$~\cite{Nachtmann:1998va, *Nachtmann:1999ys}, Higgs~\cite{BarShalom:1995jb,
Delaunay:2013npa, Bhattacharya:2014rra,Beneke:2014sba}, and
beyond-the-standard-model~\cite{Langacker:2007ur} physics.

\subsection{Asymmetries}
The simplest \emph{up-down} \tp[-] asymmetries are based on the sign of one of
the constructible \tp\ (see \autoref{eq:tp_int})
\begin{equation}
\mathsf{f}(\ps)=\sig\{\,\epsilon_{\mu\nu\rho\sigma} \; a^\mu b^\nu c^\rho d^\sigma\,\}.
	\label{eq:simplest_tp}
\end{equation}
The usual quantities defined in the literature
\begin{equation*}
\parenbar{A}_\M	\equiv
	\frac{	\ips\:\mathsf{f}(\ps)\:
	\big[\dgs{\M-odd}{\CP-even}	
			\parenpm\dgs{\M-odd}{\CP-odd}
			\big]
	}{	\ips	\big[\dgs{\M-even}{\CP-even}
			\parenpm\dgs{\M-even}{\CP-odd}\big]
	}
\end{equation*}
are ratios of integrated \M-odd and \M-even differential rates and
have no definite \CP\ transformation properties. The converse could only be
argued when differences of \CP-even phases are proven vanishing. In the
notations of \autoref{sec:phases},
\begin{equation*}
	A_\M \propto
	2\:a_e^j\:a_o^k\:
	\sin\big[ (\delta_e^j-\delta_e^k) + (\varphi_e^j-\varphi_o^k)\big]
\end{equation*}
then actually becomes a probe for small differences in \CP-odd phases
$\varphi$. On the contrary,
\begin{equation*}
	\mathcal{A}_\M^\CP	\equiv \frac{1}{2} (A_\M - \bar A_\M),
\end{equation*}
is always \CP\ odd. $\bar A_\P$ is occasionally defined as the
$\CP\M$ conjugate of $A_\M$ and has then a sign opposite to $\bar A_\M
\equiv \CP[ A_\M ]$ defined here. With this alternative convention,
$\mathcal{A}_\M^\CP$ becomes a sum. Other asymmetries were for
instance listed in Ref.~\cite{Bevan:2014nva}. Instead of
$\mathcal{A}_\M^\CP$, one may consider
\begin{equation*}
	\mathcal{\widetilde{A}}_\M^\CP 
	\equiv \frac{\ips\:\mathsf{f}(\ps)\:
		\dgs{\M-odd}{\CP-odd}}{\ips
		\dgs{\M-even}{\CP-even}}.
\end{equation*}
This choice corresponds to the more common one when the total rate asymmetry of
\autoref{eq:rate_asym} vanishes. Using $\mathcal{\widetilde{A}}_\M^\CP$, the
\M-even--\CP-odd and \M-odd--\CP-odd families of observables can be kept
independent. Uncertainties in the relative abundance of the two \CP-conjugate
initial states can however make the use of $\mathcal{A}_\M^\CP$ experimentally
preferable.

\subsection{Dilutions and \texorpdfstring{$\mathsf{f}(\ps)$}{f(Phi)} sets}
The `$\sig$' function used in \autoref{eq:simplest_tp} is not the only possible
weight function $\mathsf{f}(\ps)$ that could be used in phase-space integrals of
\M-odd differential rates. This choice amounts, experimentally, to counting
events in regions of phase space. Moreover, the adjunction of any \M-even factor
in the `$\sig$' argument besides the antisymmetric contraction
$\epsilon_{\mu\nu\rho\sigma} \; a^\mu b^\nu c^\rho d^\sigma$ would obviously
yield other potentially interesting observables. Using a basis of \M-odd
functions on $\ps$, it is also possible to decompose the \M-odd--\CP-odd
differential rate in moments (see Refs.~\cite{Dighe:1998vk, Beaujean:2015xea,
Gratrex:2015hna} about the method of moments). As in Ref.~~\cite{Aaij:2014qwa},
a binning of the phase space could also be defined and a chi-squared test
carried out to assess local departures from zero in the \M-odd--\CP-odd piece of
the differential rate. This would correspond to choosing, for the
$\mathsf{f}(\ps)$s, a set of characteristic functions that evaluate to $1$ in
one bin and vanish elsewhere. At least three categories of $\mathsf{f}(\ps)$
functions can thus be used to describe the \M-odd--\CP-odd piece of the
differential decay rate:
\begin{list}{$-$}{\leftmargin 4mm \itemsep 1mm \parsep 0mm \topsep 1mm}
\item `$\sig$' functions defining a signed partition of the phase space,
\item a \M-odd basis on $\ps$ providing a decomposition in moments,
\item characteristic functions defining a phase-space binning.
\end{list}
To avoid dilutions in the integral of \autoref{eq:tp_int}, the functions chosen
should ideally change sign wherever the \M-odd--\CP-odd piece of the
differential decay rate itself changes sign. The bins' boundaries should also be
placed there.

The question of what set of $\mathsf{f}(\ps)$ functions would yield
the best sensitivity to \CP\ violation is nontrivial and depends on the process
at hand. Actually, when the form of the differential decay rate is known with
confidence, one may rely on an unbinned likelihood fit to the data for extracting
\CP-violating parameters. Such \emph{amplitude analyses} have notably been
carried out for several $B$-meson decays: \emph{e.g.}, for $B^0_s\to
K^+K^-K^+K^-$, dominated by a $\phi\phi$ intermediate state~\cite{Aaltonen:2011rs, *Aaij:2012ud, *Aaij:2014kxa},
or for $B^0\to K^+K^-K^+\pi^-$, dominated by a $\phi K^{*0}$ resonant
intermediate state~\cite{Aubert:2008zza,Aaij:2014tpa}.

Trustworthy parametrizations also make it possible to determine the asymmetries
relevant in the study of the \CP-odd phases that might appear in perturbative
processes like $h\to \ell^+\ell^-\: \ell^{\prime+}\ell^{\prime-}$, or $e^+e^-\to
h\: \ell^+\ell^-$~\cite{Beneke:2014sba, Buchalla:2013mpa, Modak:2013sb,
Bishara:2013vya, Curtin:2013fra, Soni:1993jc}. Observables of optimal
statistical significance can then also be determined~\cite{Atwood:1991ka}.

In the hadronic decays of heavy mesons however, the parametrization provided
by a resonance model is only phenomenological and, although it may capture
accurately enough the main features of the studied process, new sources of \CP\
violation may only be observable in finer details.
Using tests of \CP\ violation that have a limited reliance upon the process
dynamics and its parametrization is therefore desirable.

\section{Spinless four-body decays}

Four-body decays involving only spinless particles are simple examples of
processes in which four independent four-vectors can be measured. In these
cases, \M\ is equivalent to \P\ and there is actually one single
independent antisymmetric $\epsilon_{\mu\nu\rho\sigma}$ contraction which
involves the external particles' four-momenta. All \M-odd functions of the phase
space are built upon it.

In the following, we will focus on this simple case and investigate how to
define appropriate signed partitions (or binnings) of the phase space. For
concreteness, we will often refer to the specific $D^0\to K^+K^-\pi^+\pi^-$
decay. Its differential rate, as well as the one of the corresponding
\CP-conjugate process, has recently been measured with an impressive accuracy by
the LHCb Collaboration~\cite{Aaij:2014qwa}. Note we will only consider
time-integrated quantities while the LHCb Collaboration also recorded the time
dependence of the decay rate.

\subsection{Phase-space parametrization}
Hadronic multibody decays often receive contributions of various topologies. The
ones so far measured in $D^0\to K^+K^-\pi^+\pi^-$ are displayed in
\autoref{tab:inter_states}.
A given resonance structure would be most appropriately described with a
parametrization of the phase space that includes the invariant masses in which
resonances occur. Such a description would likely be the most sensitive to the
interferences between the several partial-wave contributions to that topology.
On the contrary, the effects of the resonances occurring in other invariant 
masses would be diluted. Therefore, at this point already, the parametrization of
the four-body phase space challenges our aim at a description independent of the
process dynamics.

\begin{table}
\ensuremath{%
\begin{array}{@{}c@{}lc@{}}
\hline\noalign{\vskip 1mm}
	& \text{Intermediate states}
		& \Br\;\times 10^4	\\[1mm]\hline\noalign{\vskip 1mm}
\multirow{3}{*}{%
\raisebox{-5.5mm}{%
\begin{fmfgraph}(10,7)
\fmfleft{l}
\fmfright{r1,r2,r3,r4}
\fmf{vanilla,tension=10}{l,v}
\fmf{vanilla,tension=2}{v1,v,v2}
\fmf{vanilla}{r1,v1,r2}
\fmf{vanilla}{r3,v2,r4}
\end{fmfgraph}}
}
	& (\phi\rho^0)_S,\quad		\phi\to K^+K^-,\quad \rho^0\to \pi^+\pi^-
		& 9.3 \pm 1.2		\\[0mm]
	& (\phi\rho^0)_D,\quad		\phi\to K^+K^-,\quad \rho^0\to \pi^+\pi^-
		& 0.83\pm 0.23		\\[0mm]
	& (K^{*0}\overline{K}^{*0})_S,\quad	K^{*0}\to K^\pm\pi^\mp
		& 1.48\pm 0.30		\\[3mm]
\raisebox{-3.5mm}{%
\begin{fmfgraph}(10,7)
\fmfleft{l}
\fmfright{r1,r2,r3,r4}
\fmf{vanilla,tension=10}{l,v}
\fmf{vanilla,tension=2}{v,v2}
\fmf{vanilla}{r1,v,r2}
\fmf{vanilla}{r3,v2,r4}
\end{fmfgraph}
}
	& \phi (\pi^+\pi^-)_S,\quad	\phi\to K^+K^-
		& 2.50\pm 0.33		\\[6mm]
\raisebox{-3.5mm}{%
\begin{fmfgraph}(10,7)
\fmfleft{l}
\fmfright{r1,r2,r3,r4}
\fmf{vanilla,tension=10}{l,v}
\fmf{vanilla}{r1,v,r2}
\fmf{vanilla}{r3,v,r4}
\end{fmfgraph}
}
	& (K^-\pi^+)_P (K^+\pi^-)_S
		& 2.6 \pm 0.5		\\[4mm]
\multirow{6}{*}{%
\raisebox{-10.5mm}{%
\begin{fmfgraph}(10,7)
\fmfleft{l}
\fmfright{r1,r2,r3,r4}
\fmf{vanilla,tension=10}{l,v}
\fmf{vanilla}{v,r4}
\fmf{vanilla,tension=4}{v,v1}
\fmf{vanilla,tension=2}{r3,v1,v2}
\fmf{vanilla}{r1,v2,r2}
\end{fmfgraph}}
}
	& K_1^+K^-,\quad	K_1^+\to K^{*0}\pi^+
		& 1.8 \pm 0.5		\\[-.5mm]
	& K_1^-K^+,\quad	K_1^-\to \overline{K}^{*0}\pi^-
		& 0.22 \pm 0.12		\\[2mm]
	& K_1^+K^-,\quad	K_1^+\to \rho^0K^+
		& 1.14\pm 0.26		\\[-.5mm]
	& K_1^-K^+,\quad	K_1^-\to \rho^0K^-
		& 1.46\pm 0.25		\\[2mm]
	& K^*(1410)^+K^-,\;	K^*(1410)^+\to K^{*0}\pi^+
		& 1.02\pm 0.26		\\[-.5mm]
	& K^*(1410)^-K^+,\;	K^*(1410)^-\to \overline{K}^{*0}\pi^-
		& 1.14\pm 0.25
\\[1mm]\hline
\end{array}%
}%
\caption{The different measured contributions~\cite{Artuso:2012df} to
$\Br(D^0\to K^+K^-\pi^+\pi^-)=(24.3\pm1.2)\times 10^{-4}$ as
listed by the Particle Data Group~\cite{Agashe:2014kda}, the corresponding decay
topologies---or resonance structures---, and branching fractions. The $S$,
$P$, and $D$ indices indicate the partial waves in which the particle pairs are produced.}
\label{tab:inter_states}
\end{table}%

\begin{figure}
\includegraphics{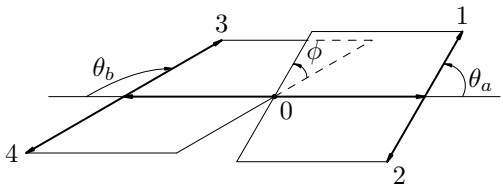}
\caption{Parametrization of the phase space of a $0\to 1\,2\,3\,4$ four-body
decay privileging the $(1\,2)$ and $(3\,4)$ subsystems. The momenta of the
final-state particle pairs are pictured in their joint rest frames.}
\label{fig:ps}
\end{figure}%

One may first consider the partial-wave decomposition of the dominant resonant
intermediate state. Let us here focus on the $0\to a\,b\to (1\,2)\,(3\,4)$
topology found in the $\phi\rho^0$ resonant contribution that accounts for about
$40\%$ of the $D^0\to K^+K^-\pi^+\pi^-$ branching fraction. Repeating the
analysis that follows for different parametrizations would be required to obtain
better sensitivities to other decay topologies. A $(1\,4)\,(2\,3)$ pairing would
for instance allow to better probe \CP\ violation involving a
$K^{*0}\overline{K}^{*0}$ resonant intermediate state in the $D^0\to
K^+K^-\pi^+\pi^-$ decay.

The standard Cabibbo-Maksymowicz parametrization~\cite{Cabibbo:1965zzb} of the
phase space $\ps$ can be adopted to describe a four-body decay of $0\to a\,b\to
(1\,2)\,(3\,4)$ topology. It is based on two invariant masses $m^2_a$ and
$m^2_b$ (which become constants in the narrow-width approximation) and three
angles (see \autoref{fig:ps}). In the $a$ and $b$ subsystems' rest frames, the
orientations of the final-state particles' momenta are respectively characterized
by $\theta_a$ and $\theta_b$, comprised in the $[0,\pi]$ interval. The relative
orientation of the planes formed by the two pairs of momenta is measured by
$\phi\in[-\pi,\pi]$. Note that $\theta_a$ and $\theta_b$ are \M-even while
$\phi$ is \M-odd (and \P-odd). The whole $\phi$ dependence of a differential
distribution of definite \M\ transformation properties can thus be obtained from
the $[0,\pi]$ interval. The angle $\phi$ also determines the sign of the \tp:
\begin{equation*}
	\epsilon_{\mu\nu\rho\sigma} \;
		p_1^\mu		\,
		p_2^\nu		\,
		p_3^\rho	\,
		p_4^\sigma
	= \frac{1}{8}m_a m_b \sqrt{\lambda(m_0^2,m_a^2,m_b^2)}\; \st1\st2 \sp.
\end{equation*}
where $\lambda(x^2,y^2,z^2)\equiv(x+y+z)(x+y-z)(x-y+z)(x-y-z)$ is the usual
Källén function, and $m_{1,2,3,4}$ have been neglected. We will occasionally use
shorthands like $\cp\equiv\cos\phi$, $\st[^2]{}\equiv\sin^2\theta$,
$\st[2]{}\equiv\sin(2\theta)$.

\subsection{Differential decay rates}
For a decay to four spinless particles forming two intermediates states of
angular momentum $j_a$, $j_b$, the amplitude can be expressed in terms of
spherical harmonics. With a spinless initial state, the two intermediate states
have equal helicities $\lambda$. We can therefore write
\begin{equation*}
\mathcal{M}=
	4\pi\;
	\sum_{j_a,j_b,\lambda}
	A^{j_a,j_b}_\lambda(m_a^2,m_b^2)	\;
	Y_{j_a}^\lambda(\theta_a,\phi)		\;
	Y_{j_b}^{\lambda}(\theta_b,0)^*
\end{equation*}
with $|\lambda|\le\min(j_a,j_b)$, and partial-wave amplitudes
$A^{j_a,j_b}_\lambda$ of mass dimension $-1$. General expressions for $n$-body
phase spaces and arbitrary spins can be derived from Refs.~\cite{Jacob:1959at,
Werle:1963gba, *Werle:1963jba, *Werle:1963lba, Berman:1965gi}. Our normalization
is chosen such that the squared amplitude integrated over the $\theta_{a,b}$ and
$\phi$ angles takes the form
\begin{equation*}
\int\!\!\frac{\d\ct a}{2}\frac{\d\ct b}{2}\frac{\d\phi}{2\pi}\; |\mathcal{M}|^2 = 
	\sum_{j_a,j_b,\lambda} |A^{j_a,j_b}_\lambda(m_a^2,m_b^2)|^2.
\end{equation*}

%
%
In the $j_a=1=j_b$ case relevant for the $\phi\rho^0$ intermediate state of
$D^0\to K^+K^-\pi^+\pi^-$, one can define the linear polarization amplitudes
\begin{equation*}
A_0	\equiv A_0^{1,1},\qquad
A_{\pa,\pe}	\equiv \frac{1}{\sqrt{2}} \left(A_{+1}^{1,1} \pm A_{-1}^{1,1} \right),
\end{equation*}
where, for conciseness, we omitted the $m_{a,b}^2$ dependences. The
amplitude can then be written as
\begin{equation*}
\frac{1}{3}\mathcal{M} =
	          A_0              \ct a\ct b
	+   \frac{A_\pa}{\sqrt{2}} \st a\st b\cp
	- i \frac{A_\pe}{\sqrt{2}} \st a\st b\sp
\end{equation*}
where the last term is \M\ odd because of its $\sp$ dependence. Its factor of
$i$ respects the phase conventions of \autoref{eq:ifactor}.
Denoting by $\bar A_{0,\pa,\pe}$ the linear polarization amplitudes of the
\CP-conjugate process, the corresponding differential decay rate can be
decomposed, as described before, into four pieces of definite \M\ and \CP\
transformation properties:
\begin{align*}
\frac{2m_0}{9}%
\dgs{\M-even}{\CP-even}
	&= \frac{|A_0|^2+|\bar A_0|^2}{2}
		\:\ct[^2]1\ct[^2]2
\\	&+ \frac{|A_\pa|^2+|\bar A_\pa|^2}{4}
		\:\st[^2]1\st[^2]2\cp[^2]
\\	&+ \frac{|A_\pe|^2+|\bar A_\pe|^2}{4}
		\:\st[^2]1\st[^2]2\sp[^2]
\\	&+ \frac{\Re\{ A_0A_\pa^* + \bar A_0\bar A_\pa^*\}}{4\sqrt{2}}
		\:\st[2]1\st[2]2\cp
,\\
\frac{2m_0}{9}%
\dgs{\M-odd}{\CP-even}
	&= \frac{\Im\{ A_\pe A_0^* + \bar A_\pe\bar A_0^*\}}{4\sqrt{2}} 
		\:\st[2]1\st[2]2\sp
\\
	&+ \frac{\Im\{ A_\pe A_\pa^* + \bar A_\pe \bar A_\pa^*\}}{4} 
		\:\st[^2]1\st[^2]2\sp[2]
,\\
\frac{2m_0}{9}%
\dgs{\M-even}{\CP-odd}
	&= \frac{|A_0|^2-|\bar A_0|^2}{2}
		\:\ct[^2]1\ct[^2]2
\\	&+  \frac{|A_\pa|^2-|\bar A_\pa|^2}{4}
		\:\st[^2]1\st[^2]2\cp[2]
\\	&+  \frac{|A_\pe|^2-|\bar A_\pe|^2}{4}
		\:\st[^2]1\st[^2]2\sp[2]
\\	&+  \frac{\Re\{ A_0A_\pa^* - \bar A_0\bar A_\pa^*\}}{4\sqrt{2}} 
		\:\st[2]1\st[2]2\cp
,\\
\frac{2m_0}{9}%
\dgs{\M-odd}{\CP-odd}
	&= \frac{\Im\{ A_\pe A_0^* - \bar A_\pe\bar A_0^*\}}{4\sqrt{2}} 
		\:\st[2]1\st[2]2\sp
\\
	&+ \frac{\Im\{ A_\pe A_\pa^* - \bar A_\pe \bar A_\pa^*\}}{4} 
		\:\st[^2]1\st[^2]2\sp[2]
,
\end{align*}
where 
\begin{align*}
\d\ps &\equiv
	\frac{\sqrt{\lambda(m_0^2,m_a^2,m_b^2)}}{8\pi m_0^2}
	\frac{\sqrt{\lambda(m_a^2,m_1^2,m_2^2)}}{8\pi m_a^2}\\
	&\frac{\sqrt{\lambda(m_b^2,m_3^2,m_4^2)}}{8\pi m_b^2}	\;
	\frac{\d\ct a}{2}	\;
	\frac{\d\ct b}{2}	\;
	\frac{\d\phi}{2\pi}	\;
	\frac{\d m^2_a}{2\pi}	\;
	\frac{\d m^2_b}{2\pi}.
\end{align*}
The linear polarization amplitudes may receive different contributions each
having a \CP-even phase $\delta^i_X$ and a \CP-odd phase $\varphi^i_X$:
\begin{equation*}
	A_X(m_a^2,m_b^2)\equiv \sum_i a^i_X(m_a^2,m_b^2)	\;
	e^{i[\delta^i_X(m_a^2,m_b^2) +\varphi^i_X]}
\end{equation*}
for real-valued $a^i_X$, $\delta^i_X$, $\varphi^i_X$ and $X=0$, $\pa$, or $\pe$.
The corresponding \CP-conjugate quantities are then given by $\bar A_X\equiv \sum_i a^i_X
\: e^{i[\delta^i_X -\varphi^i_X]}$, so that
\begin{multline*}
\Re\{A_XA_Y^* \pm\bar A_X\bar A_Y^*\}/2
	\\\shoveright{
	=
		\pm\sum_{i,j} a^i_X a^j_Y\;\;
		{}^\text{cos}_\text{sin}(\delta^i_X-\delta^j_Y)\;\;
		{}^\text{cos}_\text{sin}(\varphi^i_X-\varphi^j_Y)
,}\\
\shoveleft{\Im\{A_XA_Y^* \pm\bar A_X\bar A_Y^*\}}/2
	\\= +\sum_{i,j} a^i_X a^j_Y\;\;
		{}^\text{sin}_\text{cos}(\delta^i_X-\delta^j_Y)\;\;
		{}^\text{cos}_\text{sin}(\varphi^i_X-\varphi^j_Y)
.
\end{multline*}
In this specific example, again, the different pieces of the partial rate
exhibit the sensitivities to the \CP-even and -odd phases described earlier.

\subsection{Beyond the most common observables}
Interestingly, with a single resonant intermediate state having $j_a=1=j_b$, the
total rate asymmetry based on the integral of \autoref{eq:rate_asym}
vanishes when the $A_{0}$ coefficient receives contributions of
identical phases, or one single contribution. (The terms involving other linear
polarization amplitudes vanish upon phase-space integration.) In such a case,
only could a differential rate study provide information about \CP\ violation.

Without assumption about the presence of identical phases, the most
common \emph{up-down} integrated asymmetry based on the sign of the \tp
\begin{equation*}
	\ips\:\sig\{\sp\}\:\dgs{\M-odd}{\CP-odd}	
\end{equation*}
also vanishes in this simple case.
This illustrates---in an extreme way---that
phase-space integration may result in losses of sensitivity to
\CP-violating phases. A nontrivial phase-space-dependent \M-even factor in
the \M-odd--\CP-odd differential rate can make it change sign where the \tp\
does not.

Such dilutions can obviously be overcome when a trustworthy
parametrization of the differential rate is known. Taking seriously the
simplified parametrization of the $D^0$ decay presented above, the bare
examination of the differential rates indicates that more information about \CP-odd
and -even phases is contained in the piecewise integrals of
\autoref{tab:more_asym} upon which asymmetries could be constructed.

However, as already stressed, the parametrization of heavy mesons' hadronic
decays is only phenomenological and may miss some fine interference details that
have the potential of revealing new sources of \CP\ violation. We would
therefore wish to adopt a more systematic approach that does not rely on strong
theoretical assumptions about the process dynamics.

\begin{figure*}
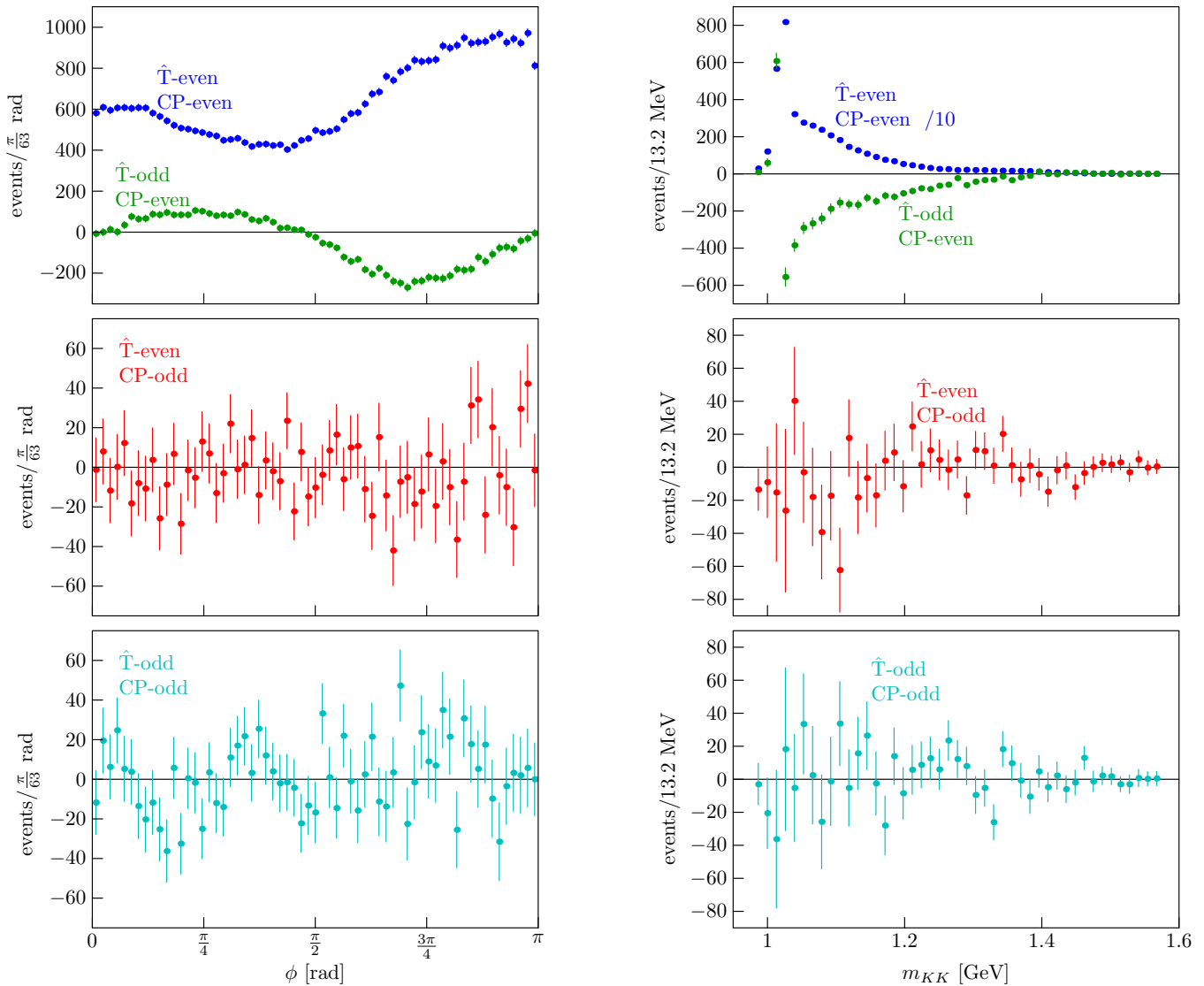

\includegraphics[scale=\myscale]{cosph}\hfill
\includegraphics[scale=\myscale]{mkk}
\caption{Decomposition of the measured $D\to K^+K^-\pi^+\pi^-$ differential rate
into components of definite \M\ and \CP\ transformation properties, projected
onto the $\phi$ angle and $m_{KK}$ invariant mass. The uncertainties on the LHCb
data points in Fig.~3(e-f) and Fig.~2(c-d) of Ref.~\cite{Aaij:2014qwa} have been
assumed equal to $\sqrt{N}+8$ and uncorrelated.}
%
%
\label{fig:cosph}
\label{fig:mkk}
\end{figure*}

\subsection{A first look at the data}
This point can be made more concrete using the recent experimental study of the
$D^0\to K^+K^-\pi^+\pi^-$ decay. The LHCb Collaboration displayed in
Ref.~\cite{Aaij:2014qwa} the measured $m_{\pi^+\pi^-}$, $m_{K^+K^-}$,
$\cos\theta_\pi$, $\cos\theta_K$ and $\phi$ distributions for both $D^0$ and
$\bar D^0$ as well as $\sin\phi>0$ and $\sin\phi<0$. This allows for the
marginalized differential distributions of definite \M\ and \CP\ properties to
be derived.
The left panel of \autoref{fig:cosph} for instance shows those differential
distributions projected onto the $\phi$ angle (\emph{i.e.}, marginalized over
the four other phase-space variables). The respective $\cp[^2]$ and $\sp[2]$
dependences of the \M-even--\CP-even and \M-odd--\CP-even differential rates
expected from a dominant $\phi\rho^0$ contribution are clearly visible while the
\M-even--\CP--odd and \M-odd--\CP-odd distributions are roughly compatible with
zero.

An oscillatory pattern can however be distinguished in the \M-odd--\CP-odd
differential rate. The $A_n\equiv\ips \:\sig\{\sin n\phi\}\:
\dgs{\M-odd}{\CP-odd}$ asymmetries (see \autoref{tab:phi_asym}) notably point at
the presence of a sizable $\sin8\phi$ contribution: the $A_8$ departure from
zero is of about $2.6$ standard deviations ($2.0$ standard deviations for $A_2$
and $A_{13}$).
If genuine, this rapid oscillatory behavior would indicate the presence of a
\CP-violating phase difference but would not have contributed to the asymmetries
that could be expected from a simple $\phi\rho^0$ parametrization.
Whether any resonance model considered as providing a fair description of that
process would have included a contribution oscillating so rapidly is also
unclear.

\begin{table*}
\ensuremath{%
\begin{aligned}
\hline\noalign{\vskip 3mm}
\ips	\sig\{\ct a\ct b\cp\}	\dgs{\M-even}{\CP-even}
	=& +
		\frac{2\sqrt{2}}{\pi}
		\int\!\!\frac{\d m^2_a}{2\pi}\frac{\d m^2_b}{2\pi}\; \mathcal{N}\;
		\sum_{i,j} a^i_0\: a^j_\pa\;
		\cos(\delta^i_0-\delta^j_\pa)\;	\cos(\varphi_0^i-\varphi_\pa^j)
\\
\ips	\sig\{\ct a\ct b\sp\}	\dgs{\M-odd}{\CP-even}
	=& +
		\frac{2\sqrt{2}}{\pi}
		\int\!\!\frac{\d m^2_a}{2\pi}\frac{\d m^2_b}{2\pi}\; \mathcal{N}\;
		\sum_{i,j} a^i_\pe\: a^j_0\;
		\sin(\delta^i_\pe-\delta^j_0)\;	\cos(\varphi_\pe^i-\varphi_0^j)
\\
\ips	\sig\{\ct a\ct b\cp\}	\dgs{\M-even}{\CP-odd}
	=& -
		\frac{2\sqrt{2}}{\pi}
		\int\!\!\frac{\d m^2_a}{2\pi}\frac{\d m^2_b}{2\pi}\; \mathcal{N}\;
		\sum_{i,j} a^i_0\: a^j_\pa\;
		\sin(\delta^i_0-\delta^j_\pa)\;	\sin(\varphi_0^i-\varphi_\pa^j)
\\
\ips	\sig\{\ct a\ct b\sp\}	\dgs{\M-odd}{\CP-odd}
	=& +
		\frac{2\sqrt{2}}{\pi}
		\int\!\!\frac{\d m^2_a}{2\pi}\frac{\d m^2_b}{2\pi}\; \mathcal{N}\;
		\sum_{i,j} a^i_\pe\: a^j_0\;
		\cos(\delta^i_\pe-\delta^j_0)\;	\sin(\varphi_\pe^i-\varphi_0^j)
\\
\ips	\sig\{\sp[2]\}	\dgs{\M-odd}{\CP-even}
	=& +
		\frac{4}{\pi}
		\int\!\!\frac{\d m^2_a}{2\pi}\frac{\d m^2_b}{2\pi}\; \mathcal{N}\;
		\sum_{i,j} a^i_\pe\: a^j_\pa\;
		\sin(\delta^i_\pe-\delta^j_\pa)\;	\cos(\varphi_\pe^i-\varphi_\pa^j)
\\
\ips	\sig\{\sp[2]\}	\dgs{\M-even}{\CP-odd}
	=& -
		\frac{4}{\pi}
		\int\!\!\frac{\d m^2_a}{2\pi}\frac{\d m^2_b}{2\pi}\; \mathcal{N}\;
		\sum_{i,j} a^i_\pe\: a^j_\pe\;
		\sin(\delta^i_\pe-\delta^j_\pe)\;	\sin(\varphi_\pe^i-\varphi_\pe^j)
\\
\ips	\sig\{\sp[2]\}	\dgs{\M-odd}{\CP-odd}
	=& +
		\frac{4}{\pi}
		\int\!\!\frac{\d m^2_a}{2\pi}\frac{\d m^2_b}{2\pi}\; \mathcal{N}\;
		\sum_{i,j} a^i_\pe\: a^j_\pa\;
		\cos(\delta^i_\pe-\delta^j_\pa)\;	\sin(\varphi_\pe^i-\varphi_\pa^j)
\\
\ips	\sig\{\cp[2]\}	\dgs{\M-even}{\CP-odd}
	=& -
		\frac{4}{\pi}
		\int\!\!\frac{\d m^2_a}{2\pi}\frac{\d m^2_b}{2\pi}\; \mathcal{N}\;
		\sum_{i,j} a^i_\pa\: a^j_\pa\;
		\sin(\delta^i_\pa-\delta^j_\pa)\;	\sin(\varphi_\pa^i-\varphi_\pa^j)
\\
\text{with }\mathcal{N} \equiv&
	\frac{1}{2m_0}
	\frac{\sqrt{\lambda(m_0^2,m_a^2,m_b^2)}}{8\pi m_0^2}
	\frac{\sqrt{\lambda(m_a^2,m_1^2,m_2^2)}}{8\pi m_a^2}
	\frac{\sqrt{\lambda(m_b^2,m_3^2,m_4^2)}}{8\pi m_b^2}
\\[3mm]\hline
\end{aligned}
}
\caption{Piecewise integrals from which information about the
\CP-conserving and \CP-violating phases between different polarization
amplitudes could be extracted, for a $0\to(1\,2)(3\,4)$ decay involving spinless
particles and proceeding through two intermediate vector resonances.}
\label{tab:more_asym}
\end{table*}

\begin{table}
\ensuremath{%
\begin{array}[t]{*{2}{c}}
\hline
n	&A_n	\\[1mm]\hline
1	&+58	\pm132	\\
2	&-259	\pm132	\\
3	&-2	\pm132	\\
4	&-134	\pm132	\\
5	&-225	\pm132	\\\hline
\end{array}
\qquad
\begin{array}[t]{*{2}{c}}
\hline
n	&A_n		\\[1mm]\hline
6	&+164	\pm132	\\
7	&+101	\pm132	\\
8	&+337	\pm132	\\
9	&-40	\pm132	\\
10	&+41	\pm132	\\\hline
\end{array}
\qquad
\begin{array}[t]{*{2}{c}}
\hline
n	&A_n		\\[1mm]\hline
11	&+128	\pm132	\\
12	&+164	\pm132	\\
13	&+268	\pm132	\\
14	&-107	\pm132	\\\hline
\end{array}
}
\caption{$A_n\equiv\ips \:\sig\{\sin n\phi\}\: \dgs{\M-odd}{\CP-odd}$
asymmetries in the data collected by the LHCb Collaboration on the $D^0\to
K^+K^-\pi^+\pi^-$ decay. The uncertainties on the data points of Fig.~3(e-f) in
Ref.~\cite{Aaij:2014qwa} have been assumed equal to $\sqrt{N}+8$ and
uncorrelated.}
\label{tab:phi_asym}
\end{table}

\begin{table}
\ensuremath{%
\begin{array}{@{\quad}r@{\:}l@{\quad}}
\hline
f_0(\ct{})	&= 1,	\\
f_1(\ct{})	&= \ct{},	\\
f_2(\ct{})	&= 3\ct[^2]{}-1,	\\
f_3(\ct{})	&= \ct{}(3\ct[^2]{}-1),	\\
f_4(\ct{})	&= \ct{}(5\ct[^2]{}-3),	\\
f_5(\ct{})	&= 5\ct[^2]{}-1, \\
f_6(\ct{})	&= 5\ct[^2]{}-3,	\\
f_7(\ct{})	&= \ct{}(5\ct[^2]{}-1),	\\
f_8(\ct{})	&= \ct{}(3\ct[^2]{}-1)\;(5\ct[^2]{}-3),	\\
f_9(\ct{})	&= (3\ct[^2]{}-1)\;(5\ct[^2]{}-1),	\\
f_{10}(\ct{})	&= \ct{}(5\ct[^2]{}-3)\;(5\ct[^2]{}-1),	\\
\cdots
\\\hline
\end{array}
}
\caption{Natural set of functions of the $\theta_{a,b}$ angles for the
systematic construction of asymmetries in $0\to(1\,2)(3\,4)$ decays involving
spinless particles.}
\label{tab:f_fun}
\end{table}

\begin{table}
\ensuremath{%
\begin{array}{r@{\qquad}*{3}{c}}
\hline
	&\mathsf{a}_X\,[\text{GeV}^{-1}]	&\delta_X	&\varphi_X	\\[1mm]\hline
X=\;0:	& 1		& 1		& 0	\\
\pa\::	& 2		& 0		& 0	\\
\pe\::	& 1		& 1		& 0.05	\\\hline
\end{array}%
}
\caption{Parameters chosen in the toy simulation of the $D\to
\phi\rho^0\to(K^+K^-) (\pi^+\pi^-)$ process.}
\label{tab:simu_params}
\end{table}

\subsection{Even more angular asymmetries}
Clearly, one way in which the presence of \CP-violating phases could be probed
without relying on a full description of the dynamics of the process studied
would be to evaluate systematically a wider range of \tp[-] asymmetries of the form
\begin{equation*}
	\ips	\:\sig\{ f_l(\ct a)\; f_m(\ct b)\; \sin n\phi\}
		\:\dgs{\M-odd}{\CP-odd}
\end{equation*}
for all combinations of reasonably large integers $l$, $m$, and $n$. In the case
of spinless final states forming two pairs of resonant intermediate states, the
natural set of functions $f$ are products of the various $\ct{}$
dependences arising in spherical harmonics:
\begin{equation*}
\begin{array}{r@{\quad}ccc}
\mathsf{l}=1:	& \ct{}, \\
2:	& 3\ct[^2]{}-1, \\
3:	& \ct{}(5\ct[^2]{}-3),	& 5\ct[^2]{}-1, \\
4:	& 35\ct[^4]{}-30\ct[^2]{}+3,	& \ct{}(7\ct[^2]{}-3),	& 7\ct[^2]{}-1,\\[-1mm]
\ldots
\end{array}
\end{equation*}
The dependences upon $\st{}$ have been dropped as they have no influence on the sign
of the associated Legendre polynomials $P^\mathsf{m}_\mathsf{l}$. Here, the set
of $f$ functions could therefore be defined as in \autoref{tab:f_fun}, keeping
in mind that another choice would be needed for final states carrying spin.

\subsection{Invariant mass dependence}
Let us still focus on the parametrization of the phase space privileging the
$0\to (1\,2)(3\,4)$ type of topology. Upon phase-space integration, the
$m_{a,b}^2$ invariant mass dependence of the decay rate could also lead to
losses of sensitivity to \CP-violating phases. This happens when it causes the
\M-odd--\CP-odd piece of the differential decay rate to change sign. Guessing
where this could happen is in general difficult. However, when resonances are
clearly identified, one at least knows the real parts of the associated propagators
\begin{equation*}
	\Re\left\{\frac{1}{m_a^2-M^2+i\Gamma M}\right\}
	= \frac{m_a^2-M^2}{(m_a^2-M^2)^2+\Gamma^2 M^2}
\end{equation*}
change sign at the resonances (and could possibly appear in interferences).

Once again, a glimpse at the LHCb data shows such a behavior actually occurs
in the $m_{KK}$ invariant mass spectrum, although in the \M-odd--\CP-even
piece of the differential rate which is not directly relevant for the extraction
of \CP-violating phases (see right panel of \autoref{fig:mkk}).

Therefore, when constructing asymmetries systematically one may also wish to
consider $\sig\{m_a^2-M^2_i\}$ and $\sig\{m_b^2-M^2_j\}$ as weight functions,
for the known resonances appearing at $M^2_{i,j}$ in the $m_{a,b}^2$ invariant
mass spectra.

\begin{figure*}
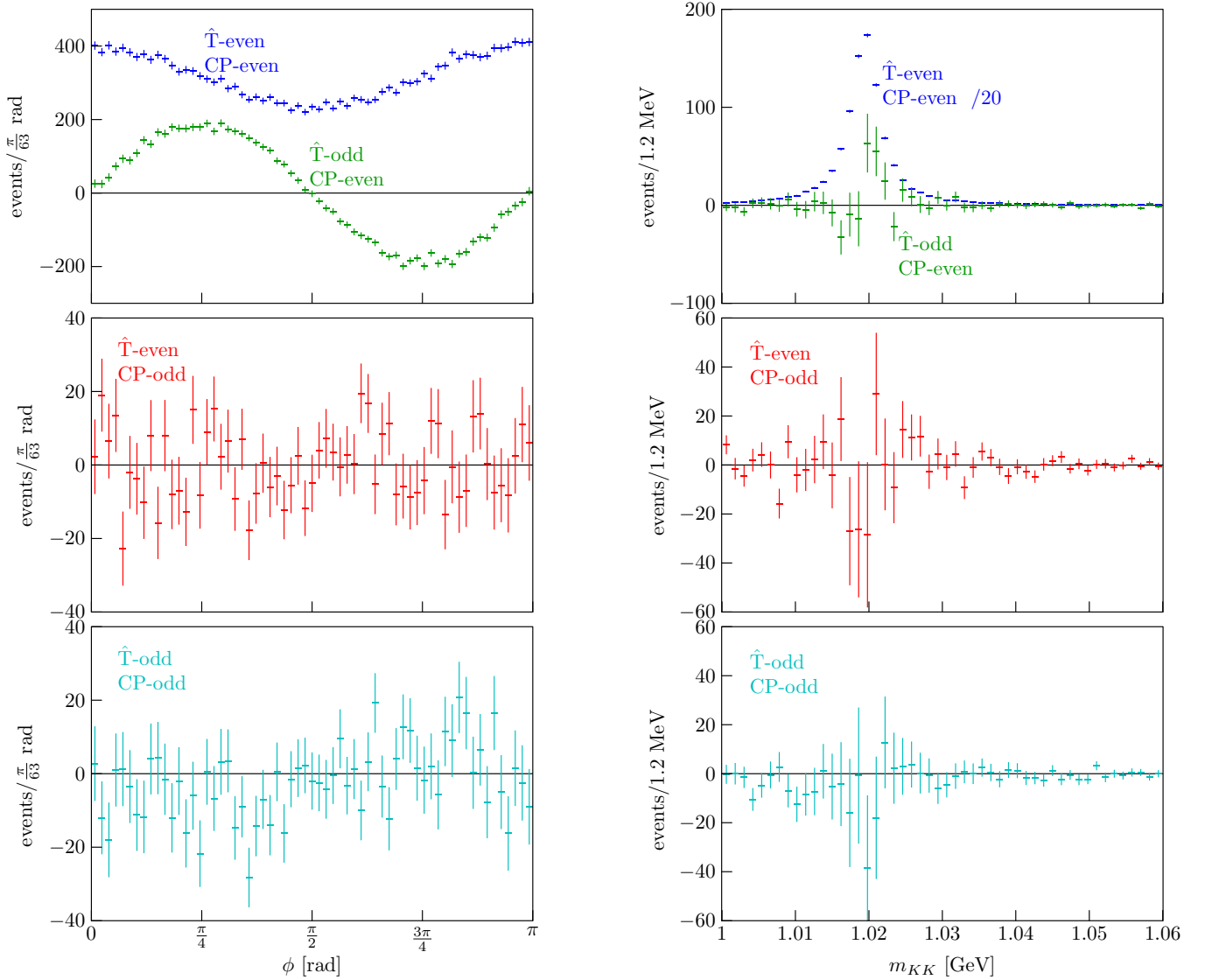

\includegraphics[scale=\myscale]{simu_phi}\hfill
\includegraphics[scale=\myscale]{simu_mkk}
\caption{Simulated $D\to \phi\rho^0\to(K^+K^-) (\pi^+\pi^-)$ decay and partial
rate decomposition in components of definite \M\ and \CP\ transformation
properties as in \autoref{eq:decomp}, projected onto the $\phi$ angle and
$m_{KK}$ invariant mass. Only statistical uncertainties are displayed.}
\label{fig:simu_phi}
\label{fig:simu_mkk}
\end{figure*}

\begin{table}[t]
\ensuremath{%
\begin{array}{rcccc}
\hline
	&\sin n\phi
		&\sin n\phi 
			&\sin n\phi 
				& \sin n\phi\\[-1.5mm]
	&	&(m_{KK}^2-m_\phi^2)
			&(m_{\pi\pi}^2-m_\rho^2)
			&\cos\theta_{KK}\cos\theta_{\pi\pi}	\\[1mm]\hline
n=1:	&-2.0	&0.86		&1.2		&-2.8	\\
2:	&-4.0	&0.88		&3.4		&0.40	\\
3:	&0.20	&0.15		&-0.014		&-1.4	\\
4:	&0.30	&-0.014		&-0.52		&-1.5	\\
5:	&0.30	&0.95		&-1.2		&-0.65	\\
6:	&-0.40	&0.20		&1.0		&0.057	\\
7:	&-2.0	&2.4		&2.4		&-0.042	\\
8:	&-0.70	&0.37		&1.5		&0.27	\\
9:	&-0.60	&-0.69		&0.88		&-0.75	\\
10:	&0.60	&-2.2		&-0.78		&1.0	\\
11:	&-2.0	&0.53		&2.1		&-0.15	\\
12:	&-0.20	&-0.092		&0.55		&-1.7	\\
13:	&-1.0	&0.30		&1.2		&0.67	\\
14:	&0.20	&1.7		&0.30		&-0.70	\\[1mm]\hline
\end{array}%
}
\caption{Departure from zero expressed in standard deviations for a few 
asymmetries, computed with the simulated sample of $D\to \phi\rho^0\to(K^+K^-)
(\pi^+\pi^-)$ decays. Only statistical uncertainties are accounted for.}
\label{tab:simu_asym}
\end{table}

\subsection{Binned analyses}
Instead of constructing asymmetries, one may rather adopt the approach of
Ref.~\cite{Aaij:2014qwa} and bin the phase space. Care must however be taken in
the binning choice. Putting together in one bin, regions of the phase
space in which the \M-odd--\CP-odd part of the differential rate changes sign
would result in sensitivity losses.

These can be assessed using a toy simulation. We
considered massless kaons and pions and generated events using
MadGraph5~\cite{Alwall:2014hca} with the following matrix elements for the
$D\rho\phi$, $\phi KK$, and $\rho\pi\pi$ interactions:
\begin{align*}
&\begin{aligned}
D\rho\phi: \epsilon_\rho^\mu\; \epsilon_\phi^\nu\; p_\rho^\alpha\; p_\phi^\beta
\;\bigg(&\mathsf{A_0}\;	g_{\alpha\nu}g_{\beta^\mu}\\[-2mm]
+\mathsf{A_\pa}	&\Big\{ g_{\mu\nu} g_{\alpha\beta}\;
	\Big[1- \dfrac{p_\rho^2\:p_\phi^2}{(p_\rho\cdot p_\phi)^2}\Big]
	- g_{\alpha\nu}g_{\beta^\mu} \Big\}\\[-2mm]
+\mathsf{A_\pe}	&\,i\epsilon_{\mu\nu\alpha\beta}
\bigg),
\end{aligned}
\\
&\phi KK: 
	\epsilon_\phi^\mu \left( p_{K^+\mu}-p_{K^-\mu}\right),
\\
&\rho\pi\pi: 
	\epsilon_\rho^\mu \left( p_{\pi^+\mu} -p_{\pi^-\mu}\right).
\end{align*}
The linear polarization amplitudes described earlier are then
\begin{small}%
\begin{align*}
A_0
	&=\frac{
			\mathsf{A_0}\:
			\lambda(m_D^2,m_{KK}^2,m_{\pi\pi}^2)
		}{
			12
			(m_{KK}^2-m_\phi^2+im_\phi \Gamma_\phi)
			(m_{\pi\pi}^2-m_\rho^2+im_\rho \Gamma_\rho)
		},
\\[2mm]
A_\pa
	&=\frac{
			\mathsf{A_\pa}\:
			\frac{m_{KK}m_{\pi\pi}}
				{m_D^2-m_{KK}^2-m_{\pi\pi}^2}
			\lambda(m_D^2,m_{KK}^2,m_{\pi\pi}^2)
		}{
			6
			(m_{KK}^2-m_\phi^2+im_\phi \Gamma_\phi)
			(m_{\pi\pi}^2-m_\rho^2+im_\rho \Gamma_\rho)
		},
\\[2mm]
A_\pe
	&=\frac{
			\mathsf{A_\pe}\:
			m_{KK}m_{\pi\pi}
			\sqrt{\lambda(m_D^2,m_{KK}^2,m_{\pi\pi}^2)}
		}{
			6
			(m_{KK}^2-m_\phi^2+im_\phi \Gamma_\phi)
			(m_{\pi\pi}^2-m_\rho^2+im_\rho \Gamma_\rho)
		}.
\end{align*}%
\end{small}%
where each of the $\mathsf{A_{0,\pa,\pe}}$ were given both a \CP-even and
\CP-odd phase: $\mathsf{A}_X = \mathsf{a}_X\: e^{i(\delta_X+\varphi_X)}$ for
$X=0,\pa,\pe$. These parameters were fixed as in \autoref{tab:simu_params} and
$40\,000$ $D^0$ and $\bar D^0$ decays generated. The decomposition of the $\phi$
differential distribution obtained is displayed in the left panel of
\autoref{fig:simu_phi}.

A larger magnitude for $\mathsf{a}_\pa$ than for $\mathsf{a}_\pe$ causes the
\M-even--\CP--even piece of the differential rate to have a dip at $\pi/2$. The
nonvanishing difference in \CP-conserving phases $\delta_\pa-\delta_\pe$
sources the $\sin2\phi$ dependence of the \M-odd--\CP-even contribution. No
structure is generated in the \M-even--\CP-odd distribution while a small
difference in \CP-violating phases $\varphi_\pa-\varphi_\pe$ allows for a
$\sin2\phi$ dependence in the \M-odd--\CP-odd differential rate. Due to limited
statistics, the latter is barely visible in \autoref{fig:simu_phi}. Additionally, a
$\sin2\theta_a\:\sin2\theta_b\sin\phi$ dependence of each piece of the
differential rate is washed out upon integration over the $\theta_{a,b}$ angles.
One can also notice a sign change in the \M-odd--\CP-even differential rate
projected on the $m_{KK}$ variable at the $m_\phi=1.02$~GeV resonance (see
right panel of \autoref{fig:simu_mkk}).

Computing
\vspace*{-3mm}
\begin{multline}
	\ips \:\sig\bigg\{
		f_l(\ct a)\; f_m(\ct b)\; \sin n\phi\\[-3mm]
		\prod_i (m_a^2-M_i^2) \prod_j (m_b^2-M_j^2)
	\bigg\} \:\dgs{\M-odd}{\CP-odd}
\label{eq:asym_prescr}
\end{multline} asymmetries as prescribed earlier, one observes the expected
excesses for $(l,m;n)=(0,0;2)$ and $(1,1;1)$. They are of $4.0$ and $2.8$
standard deviations, respectively (see \autoref{tab:simu_asym}, only statistical
uncertainties have been accounted for). Using additional
$\sig\{m_{KK}^2-m_\phi^2\}$ and $\sig\{m_{\pi\pi}^2-m_\rho^2\}$ weight functions
does not enhance the excesses' significance.

\begin{figure*}
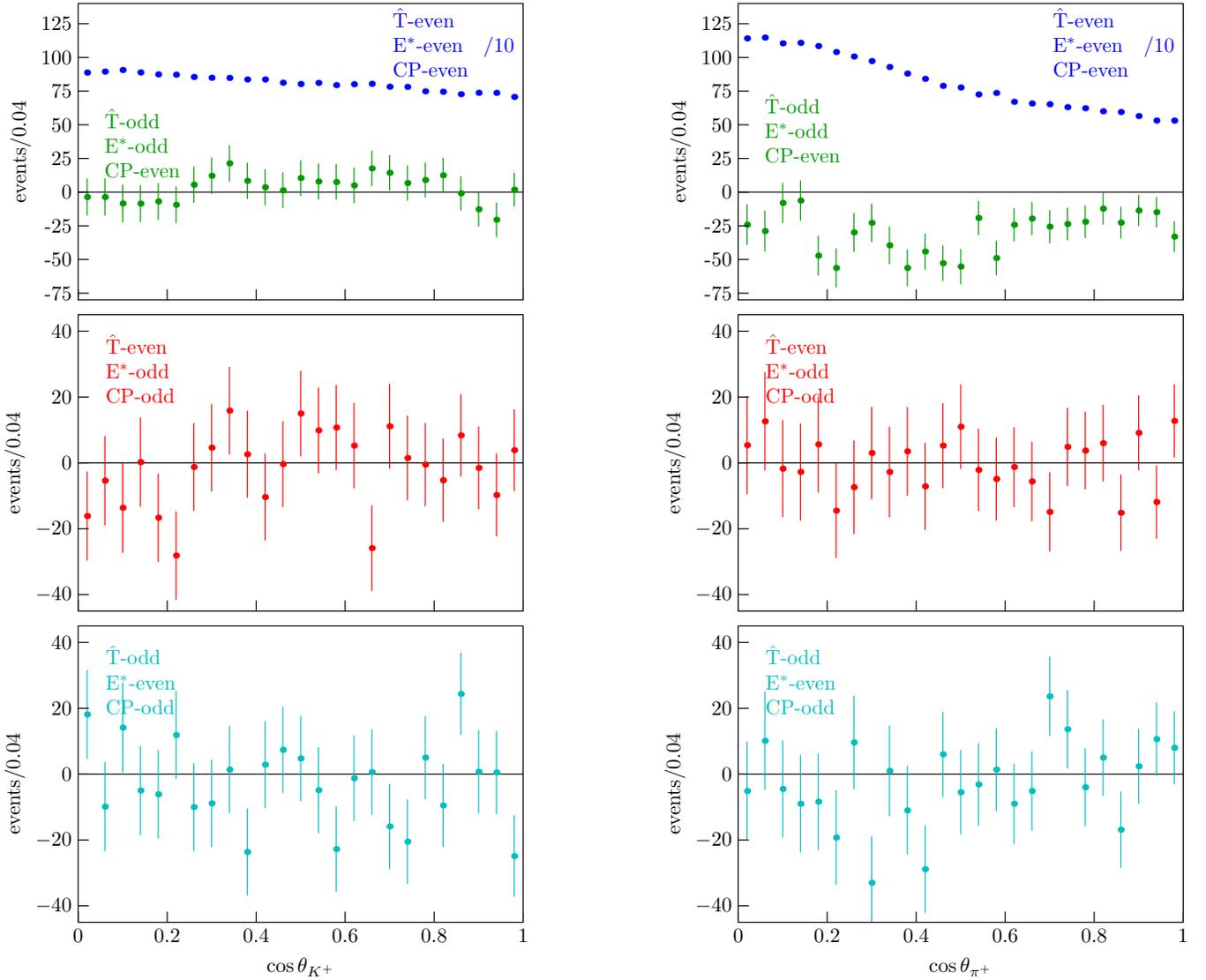

\includegraphics[scale=\myscale]{cosk}\hfill
\includegraphics[scale=\myscale]{cosp}
\caption{The four components of the $D\to K^+K^-\pi^+\pi^-$ differential rate
having definite \M, $\E^*$, and \CP\ transformation properties that could have
been measured with an untagged sample. The uncertainties on the LHCb data points
in from Fig.~3(a-d) of Ref.~\cite{Aaij:2014qwa} have been assumed equal to
$\sqrt{N}+8$ and uncorrelated.}
\label{fig:cosk}
\label{fig:cosp}
\end{figure*}

The LHCb Collaboration partitioned the phase space in $32$ bins (two bins per
kinematic variable) and estimated the combined departure from zero using a
chi-squared test~\cite{Aaij:2014qwa}. The separation between the two bins of the
$\phi$ variable was set at $1.99$~rad (its domain is restricted to the $[0,\pi]$
interval here) and between $-0.28$ and $+0.28$ for $\cos\theta_{KK}$ and
$\cos\theta_{\pi\pi}$.

In our simulated sample, a chi-squared test with only two bins of boundary located
at $\pi/2$ in the $\phi$ variable gives a departure from zero for the
\M-odd--\CP-odd differential rate equivalent to $3.9$ Gaussian standard
deviations. Once the bins' boundary is moved to $1.99$~rad, this significance
slightly diminishes to $3.7\,\sigma$. The loss of sensitivity is more significant
for the only other binning relevant to this simplified simulation. With two bins
in both the $\cos\theta_{KK}$ and $\cos\theta_{\pi\pi}$ directions, the
\M-odd--\CP-odd differential rate departs from zero at the $2.8\,\sigma$ level
when the bin boundaries are chosen at $0$, and at the $1.1\,\sigma$ level only
when they are respectively taken at the extreme values of $-0.28$ and $+0.28$.

The multiplication of unnecessary bins also leads to losses of sensitivity, in
this scheme. With eight bins having boundaries at $\pi/2$ in the $\phi$ angle and
$0$ in the $\cos\theta_{KK,\pi\pi}$ variables, one for instance obtains an
overall departure from zero of $3.5$ standard deviations.

\subsection{\texorpdfstring{Untagged $D$ and $B\to K^+K^-\pi^+\pi^-$ samples}{Untagged D and B > K+ K- pi+ pi- samples}}
\label{sec:untagged_ex}
Although a tagging of the $D^0$ has been carried out by the LHCb Collaboration
in this $D^0\to K^+K^-\pi^+\pi^-$ decay, the self-conjugate final state could
have motivated an untagged analysis. This is what was actually done in the
study of the $B_s^0$ decay to the very same final state~\cite{Aaij:2015kba}.

In both cases, the $\E^*$ permutation defined in \autoref{sec:untagged} sends
$\{K^+,K^-,\pi^+,\pi^-\}$ to $\{K^-,K^+,\pi^-,\pi^+\}$. In the parametrization
of the phase space adopted thus far, it therefore acts trivially on the
$m_{KK}$, $m_{\pi\pi}$ invariant masses, and on the $\phi$ azimuthal angle. The
cosines of the polar angles in the $K^+K^-$ and $\pi^+\pi^-$ subsystems undergo
the following transformations:
\begin{equation*}
	\begin{array}{l@{\:}l@{\:}l}
	\E^*[\cos\theta_{K^+}]	&= \cos\theta_{K^-}	&= -\cos\theta_{K^+},\\
	\E^*[\cos\theta_{\pi^+}]	&= \cos\theta_{\pi^-}	&= -\cos\theta_{\pi^+}.
	\end{array}
\end{equation*}
Practically, the untagged $\E^*$-odd distributions can therefore be obtained by
multiplying the weights of each recorded event by $\frac{1}{2}\sig\{
\cos\theta_{K^+} \cos\theta_{\pi^+} \}$ and by considering the absolute values
of both cosines as kinematic variables. The same procedure carried out with the
variable $\phi$ yields the \M-odd distributions. Using the LHCb
measurement~\cite{Aaij:2014qwa}, we display in \autoref{fig:cosk}
the projection onto the $\cos\theta_{K^+}$ and $\cos\theta_{\pi^+}$ variables of
the four differential rates that could have been measured with an untagged
sample of $D\to K^+K^-\pi^+\pi^-$ events.

In its analysis of the $B_s^0\to K^+K^-\pi^+\pi^-$ decay~\cite{Aaij:2015kba},
the LHCb Collaboration used a parametrization of the
phase space privileging resonances in the $K^+\pi^-$ and $K^-\pi^+$
invariant masses. A dominant $K^{*0}\overline{K}^{*0}$ intermediate state
motivated this choice.
In that parametrization, the permutation $\E^*$ exchanges the cosines of the polar
angles defined in the two subsystems $\ct{a}$ and $\ct{b}$, as well as their
respective invariant masses $m_a$ and $m_b$. Various $\E^*$-odd
asymmetries can therefore be constructed by using weight functions
$\mathsf{g}(\ps)$ (see \autoref{sec:int_obs}) proportional to either
$\ct{a}-\ct{b}$, or $m_a-m_b$. The $\E^*$-odd asymmetries measured in
Ref.~\cite{Aaij:2015kba} were the ones possibly appearing for $K\pi$ subsystems
forming partial waves of $j_{a,b}=0$ and $1$. The arguments presented here to
motivate the systematic use of a wider range of \M-odd--\CP-odd asymmetries
however also apply to \M-even--$\E^*$-odd--\CP-odd ones.

\section{Conclusions}

\CP\ violation in $K$ and $B$ decays has so far been observed mostly through
time-independent and time-integrated rate asymmetries. As multibody decays are
being measured with an ever increasing accuracy, it is desirable to devote more
attention to their rich differential distributions.

Taking, as an illustrative example, the $D^0\to K^+K^-\pi^+\pi^-$ decay whose
differential distribution has recently been studied by the LHCb
Collaboration~\cite{Aaij:2014qwa}, we proposed to measure a large set of
generalized \emph{triple-product} asymmetries. Their choice is guided by the
topology---or resonance structure---of the contribution under scrutiny, by the
spin of the particles involved, and by the location of the known resonances.
An illustration of the procedure and of the losses of sensitivity that may occur
with a suboptimal partition of the phase space was provided using a toy
simulation.
Such a procedure could obviously be applied to a wide range of other processes
in which \CP\ violation is searched for in differential distributions.

In charm decays, a signal of \CP\ violation would clearly point at new physics.
In $B$ decays, however, standard-model \CP\ violation is expected to be visible
in some cases. We did not investigate whether cleaner probes for physics beyond
the standard model could be constructed from differential observables. Clearly,
more theoretical studies in this direction would be necessary.

Our final point is to emphasize that more experimental studies are needed in
order to devise observables optimized for specific processes. With the new data
coming from LHCb and Belle~II, such a task is timely.

\section*{Acknowledgements}
We would like to warmly thank Archana Anandakrishnan for valuable discussions.
GD is a Research Fellow of the FNRS, Belgium, and of the Belgian American
Education Foundation, USA. The work of YG is supported in part by the U.S.\
National Science Foundation through grant PHY-0757868.

\end{fmffile}
\bibliographystyle{apsrev4-1_title}
\bibliography{references}

\end{document}